\documentclass[aps,prb,twocolumn,superscriptaddress]{revtex4-2}

\usepackage{graphicx} 
\usepackage{graphics}
\usepackage{epsfig}
\usepackage{appendix}
\usepackage{amsmath, amssymb, braket}
\usepackage{dirtytalk}
\usepackage{hyperref, listings}
\usepackage{bbold}
\usepackage{esvect, cancel}
\usepackage{svg}
\usepackage{comment}
\usepackage{float}

\begin{document}



\title{Scalable Photon-Mediated Two-Qubit Gates with Spectrally Noisy Quantum Emitters}




\author{Shreekanth S. Yuvarajan} 
\email{shreekan@buffalo.edu}
\affiliation{Department of Physics, University at Buffalo SUNY, Buffalo, New York 14260, USA}
\author{W. A. Coish}
\affiliation{Department of Physics, McGill University, 3600 Rue University, Montreal, Quebec, Canada H3A 2T8}
\author{David Hucul}
\affiliation{United States Air Force Research Laboratory, Rome, New York, 13441, USA}
\author{Herbert F. Fotso}
\email{hffotso@buffalo.edu}
\affiliation{Department of Physics, University at Buffalo SUNY, Buffalo, New York 14260, USA}

\begin{abstract}
\noindent When two quantum bits are coupled through a cavity, a two-qubit gate can be realized between them either in the near-resonant regime over a timescale established by the coupling strength of the qubits with the cavity, or in the dispersive regime, over a longer timescale established by the combination of the coupling strength and the frequency detuning between the qubits and the cavity.  When the qubits are spectrally noisy or differently detuned from the cavity, the fidelity for the operation can be drastically reduced in either case, precluding scalable realizations. 
We introduce the protocol for optimal cavity-enabled gates (POCEG) that is shown, through reliable numerical and analytical solutions, to overcome spectral differences between quantum bits and to achieve high fidelity between disparate/noisy quantum emitters. Namely,  for a cavity with low damping rate, we apply a sequence of pulses to the qubits at the frequency of the cavity while periodically modulating the coupling of the qubits to the cavity. Alternatively, in the case of a large damping rate, we operate in the dispersive regime and overcome spectral disparities by applying the pulses at a frequency far-detuned from the cavity. In both instances, we find for the quantum state transfer between the two qubits that, with a modest inter-pulse delay, the fidelity that would otherwise be strongly suppressed by the spectral mismatch of the qubits can be increased beyond 99.9\%. These protocols have the capacity to bring two-qubit gates between solid state systems across the threshold required for fault-tolerant quantum computing. 
\end{abstract}

\maketitle

\section{Introduction}
\label{sec:Introduction}    

\noindent Photon mediation is central to a variety of fundamental operations in quantum information processing (QIP). The optimization of relevant light-matter interfaces is thus critical to the realization of scalable quantum platforms. Relevant operations extend from the capacity to generate quantum entanglement between distant quantum nodes to the implementation of two-qubit gates~\cite{Kimble_QtumInternet2008, DLCZ, ChildressRepeater, AharonovichEnglundToth_NatPhot, AwschalomHansonZhou_NatPhot, SchoelkopfGirvin_Nature2008, Haroche2006_book}. Stringent requirements on the qubit properties often curtail the realization of high fidelity operations \cite{dyckovsky2012analysis, Ntrl_atoms_cavity_PRX2018}. In particular, due to spectral diffusion or to the intrinsic disparities between different qubits, including between nominally identical solid state systems, the realization of high fidelity two-qubit gates beyond the threshold required for fault-tolerant quantum computing\cite{NielsenChuang2010, Martinis2021} remains a challenge for various promising qubit modalities, including nuclear and electron spins in solid state systems such as quantum dots \cite{Burkard_semiconductor_qubits_2023, Tyryshkin_2006, fujita2017shuttling, qiao_hu_2021_superexchange}. To overcome the short-range nature of the exchange coupling between semiconductor spin qubits, photon mediation has been explored in different regimes of operation to implement two-qubit gates between different quantum dots \cite{DijkemaVandersypen_NatPhys2025, PhysRevX.12.021026, WarrenBarnesEconomouPRB2021, WarrenBarnesEconomouPRB2019}.
Indeed, effective coupling enabling the implementation of quantum gates can be achieved between pairs or ensembles of qubits coupled through a cavity~\cite{HagleyHaroche_PRL1997, FreyWallraff_PRL2012, EnglundLukin_NanoLett2010, ZhengGuo_PRL2000, YangHan_PRL2004, RosseauByrnes_PRA2014}. Furthermore, the modulation of qubit spectral properties and of the cavity coupling have been proposed as new avenues for efficient stationary-to-flying-qubit conversions \cite{li2024flying, tissot2024_flying_q_coupling_control, Burkard_coupling_modulation_2026}. Also, shaping spectral properties of quantum emitters with applied pulses has been actively investigated and has shown the capacity to enhance photon indistinguishability between originally different or noisy emitters \cite{unterguggenberger2026spectral, Fotso_noisyTLS2022, Fotso_TPI_PRB_2019, FotsoOtherPulsesJPhysB2018, FotsoDobrovitski_Absorption, FotsoEtal_PRL2016, vuckovic2020}. Similarly, protocols aiming to engineer efficient coupling between qubits or with a cavity by modulating spectral properties of the qubits either with continuous driving fields or with pulse protocols have been explored~\cite{SrinivasaTaylorPettaPRX2024, Beaudoin_2017, WarrenBarnesEconomouPRB2021}. Nevertheless, achieving scalable two-qubit gates with solid state qubits remains an outstanding challenge.\\

In this paper, we introduce the protocol for optimal cavity-enabled gate (POCEG) that is shown through reliable numerical and analytical solutions to overcome spectral differences between pairs of quantum bits and to achieve high fidelity for disparate/noisy quantum emitters. Namely, we consider two qubits at different frequencies and coupled through a cavity. For a cavity with a low damping rate, we apply a sequence of pulses to the qubits at the frequency of the cavity while periodically modulating the coupling of the qubits to the cavity. Alternatively, in the case of a cavity with a strong damping rate, we operate the system in the dispersive regime and overcome spectral disparities between the qubits by applying pulses at a frequency far-detuned from the cavity. In both instances, the protocol ensures near-unitary dynamics throughout the time-evolution of the entire \textit{Qubit 1 - Cavity - Qubit 2} system. We demonstrate that with a modest inter-pulse delay, the fidelity that would otherwise be drastically reduced by the spectral mismatch of the qubits can be increased beyond 99.9\%. In the resonant regime, the present protocol combines the effects of the periodic pulse sequence on the qubits spectral properties with the modulation of the coupling of the qubits to the cavity to ensure a constrained number of excitations in the system. Such modulation of the qubit coupling to the cavity has recently been revealed as an important avenue for scalability and can be reliably implemented, for instance, with the double quantum dot qubit coupled through a superconducting resonator~\cite{Burkard_coupling_modulation_2026, XuedongNori_PRB2012, DijkemaVandersypen_NatPhys2025, Borjans2020} . Altogether, the protocols presented here have the capacity of bringing a variety of two-qubit gates between solid state systems across the threshold required for fault-tolerant quantum computing.
 
The rest of the paper is organized as follows. In Sec.\ref{sec:Methods}, we present the model and the protocol along with a brief discussion of our numerical and analytical solutions, which are further expanded upon in the supplemental material (see \ref{sec:Appendix_Numerical_Method}). In Sec.\ref{sec:Results}, we present results characterizing the protocol, first for an ideal cavity where the protocol operates by inducing the behavior of a fully resonant system for a setup that would otherwise exhibit a low fidelity due to spectral disparities. We then consider the case of a cavity with a finite damping rate and show that for a large damping rate, the protocol can produce high fidelity by bringing the qubits into the same frequency far-detuned from the cavity. We end with our conclusion in Sec. \ref{sec:Conclusion}.

\begin{figure}[t] 
    \centering    \includegraphics[width=\columnwidth]{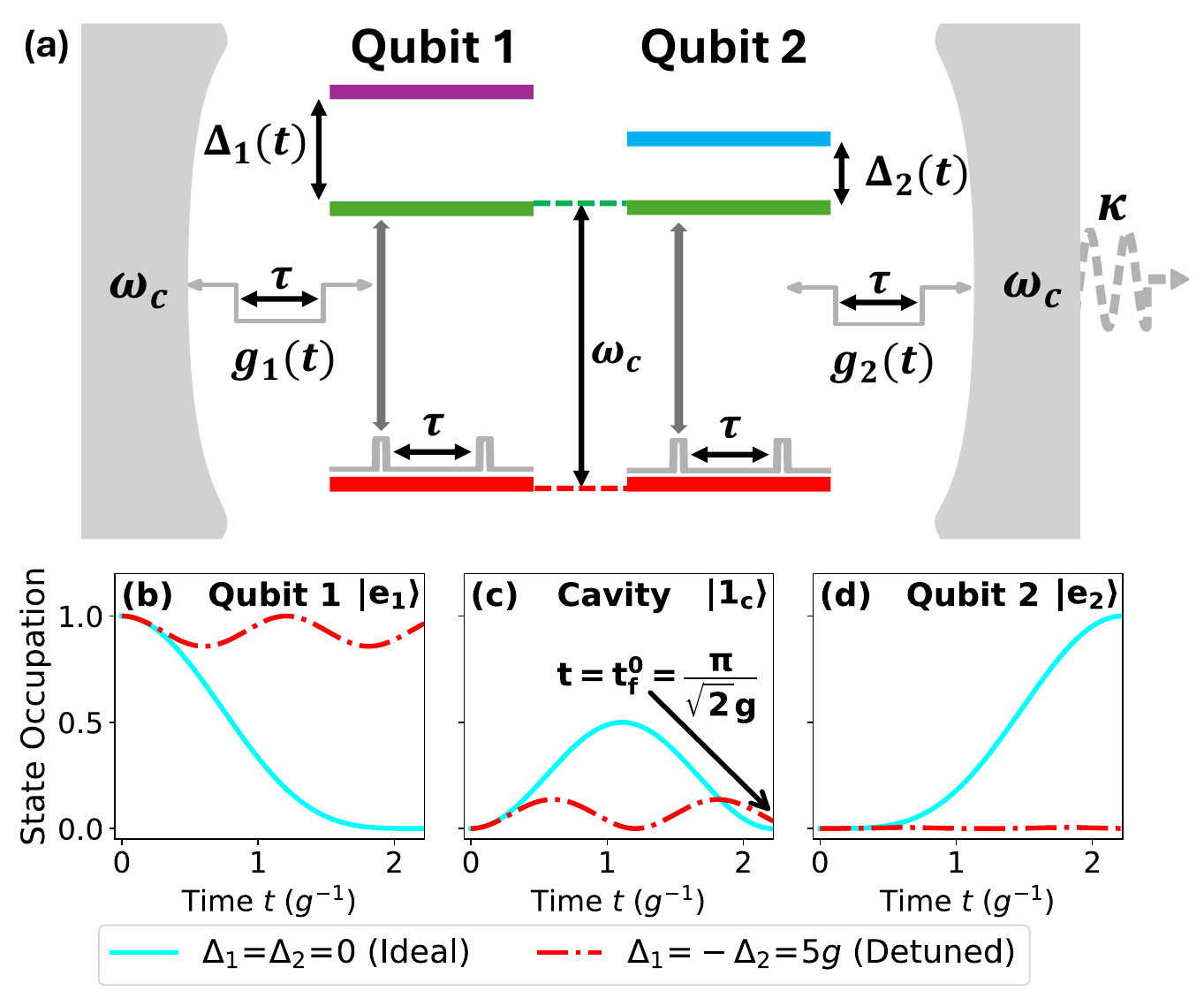}
    \caption{\textbf{(a)}: Schematic description of the POCEG applied to a system of two qubits coupled through a cavity. \textit{Qubit 1} and \textit{Qubit 2} operate at frequencies $\omega_1(t) = \omega_c + \Delta_1(t) $ and $\omega_2 (t) = \omega_c + \Delta_2(t) $ respectively and are coupled to the cavity with respective coupling strengths $g_1(t)$ and $g_2(t)$. $\Delta_1(t)$ and $\Delta_2(t)$ are the detunings of \textit{Qubit 1} and \textit{Qubit 2} respectively with the cavity that has a damping rate $\kappa$ and is at frequency $\omega_c$. 
    \textbf{(b, c, d)}: Evolution of the occupation of the excited state of the qubits and the first excited state of the cavity as a function of time without the POCEG, with $g_1=g_2=g$, and with the excitation initially in \textit{Qubit 1}, when the qubits are resonant with the cavity (cyan lines) and when the qubits are oppositely detuned from the cavity by $5g$ (red lines).
}
    \label{fig:system}
\end{figure}

\section{Model and Methods}
\label{sec:Methods}

\noindent We consider a system of two qubits coupled to a common cavity as described schematically by Fig.~\ref{fig:system} (a). \textit{Qubit 1} and \textit{Qubit 2} operate at frequencies $\omega_1 = \omega_c + \Delta_1 $ and $\omega_2 = \omega_c + \Delta_2 $ respectively and are coupled to the cavity with coupling strengths $g_1$ and $g_2$. 
$\Delta_1$ and $\Delta_2$ are thus, respectively, the detunings of \textit{Qubit 1} and \textit{Qubit 2} with the cavity that has a damping rate $\kappa$ and is at frequency $\omega_c$. The system is subjected to a control protocol involving periodic pulses and cavity modulation as discussed below.
Our system is described by the Hamiltonian:

\begin{eqnarray}
    H (t) &=&  \sum_{i=1,2} \bigg[ \frac{\hbar \omega_i}{2}\sigma^z_i + \hbar \omega_c a^{\dagger}_c a_c + \Omega(t) \sigma^x_i \nonumber \\ 
    &\quad& + \hbar g_i(t) (a_c^\dagger \sigma^-_i + a_c \sigma^+_i + a_c^\dagger \sigma^+_i + a_c \sigma^-_i) \bigg]
\label{eq:Hamiltonian_POCEG}
\end{eqnarray}

\noindent Here, $a^{\dagger}_c$ and $a_c$ are respectively the creation and annihilation operators for the cavity mode at the frequency $\omega_c$. $\sigma^z$, $\sigma^-_i = |g_i\rangle\langle e_i |$ and $\sigma^+=(\sigma^-)^\dagger$ are, respectively, the $z$-axis Pauli matrix, the lowering, and the raising operators for the two-level system, with $|g_i\rangle$ and $ |e_i\rangle$ the ground and the excited states of \textit{Qubit i}. $\Omega(t)$ is the Rabi frequency of the pulse driving field applied to the qubits by the protocol. In the absence of the control protocol, $\Omega(t) = 0$ and $g_i(t)= g_i$ is constant and we recover the two-qubit quantum Rabi Hamiltonian~\cite{QuantumRabi2}. In what follows we will use $g_1= g_2 = g$ which is used as the unit of energy or inverse time throughout this paper.

In general, the state of the system at time $t$ can be written as:
\begin{equation}
    |\Psi(t)\rangle = \sum_{\sigma_1,\sigma_2,n_c} \alpha_{\sigma_1 n_c \sigma_2}(t)|\sigma_1\rangle \otimes |n_c\rangle \otimes|\sigma_2\rangle. 
\end{equation}
Here, $\alpha_{\sigma_1 n_c \sigma_2}(t)$ is a complex coefficient, $\sigma = e$ or $g$ for Qubit $i=1,2$, and   
$n_c$ is the cavity occupation number, $n_c = 0, \; 1, \; 2, \;\cdots$. While the excitation non-conserving terms in the Hamiltonian ($\sim a_c^\dagger\sigma_i^+$, $\sim a_c\sigma_i^-$) allow for the population of higher cavity occupation numbers, as discussed in the supplemental material (\ref{sec:Appendix_Cavity_Cutoff}), the occupation of $n_c>3$ is negligible for moderate couplings. Therefore, in the results presented here, $n_c$ is capped at $3$.  
Each of the states $|\sigma_j\rangle,\,|n_c\rangle$, with $j = 1,\; 2$, lives in a two-dimensional space. So, the state of the overall system lives in an 16-dimensional space. Thus, even without the control protocol, the dynamics of the system is not broadly tractable analytically, although analytical solutions can be obtained for special system parameters and constraints \cite{yuvarajan2025cavity}. A general characterization of the dynamics of the system has been carried out using a combination of analytical and numerical methods. 
We aim to transfer the state $| \Phi_1 \rangle=a |g_1\rangle + b |e_1 \rangle$ of \textit{Qubit 1} at time $t = 0$ onto \textit{Qubit 2}. Thus, the fidelity of the operation after time $t > 0$ is $F(t) = |\langle \Psi_{\mathrm{target}}|\Psi(t)\rangle|^2$ with $|\Psi(0) \rangle=|\Phi_1\rangle\otimes |0_c\rangle \otimes |g_2\rangle$ and $|\Psi_{\mathrm{target}} \rangle=| g_1\rangle\otimes| 0_c\rangle \otimes|\Phi_2 \rangle$, where $|\Phi_i\rangle=a |g_i\rangle + b |e_i \rangle$ with the same coefficients $a,b$ for either \textit{Qubit} $i=1$ or $i=2$.

We will focus the main part of our discussion on the case of \textit{Qubit 1} initialized in the excited state ($a=0,\,b=1$),  \textit{Qubit 2} in the ground state, and the cavity initially empty. The case of the general superposition state $(a\ne 0,\,b\ne 0)$ will be discussed in \ref{sec:Appendix_Superposition_transfer}. We will also assume an ideal cavity in the first part of the results (with a vanishing cavity damping rate, $\kappa = 0$) before discussing the finite $\kappa$ case in the later part. In all solutions presented in this work, \textit{Qubit 2} is initially in the ground state. 

Fig.~\ref{fig:system} (b), (c) and (d) show, for the uncontrolled system, the time evolution between $t = 0$ and $t = t_f^0 = \pi/\sqrt{2} g$ of the excited states of \textit{Qubit 1}, the first excited state of the cavity and the excited state of \textit{Qubit 2} for a fully resonant system, $\Delta_1 = \Delta_2 = 0$ (solid cyan lines), and an oppositely detuned system, $\Delta_1 = - \Delta_2 = 5 g$ (dashed red lines). $t_f^0$ is the exact time for state transfer from \textit{Qubit 1} to \textit{Qubit 2} when both are resonant with the cavity. This time evolution illustrates the deterioration of the fidelity for qubits detuned from the cavity and/or from each other.

\begin{figure}[t] 
    \centering
    \includegraphics[width=\columnwidth]{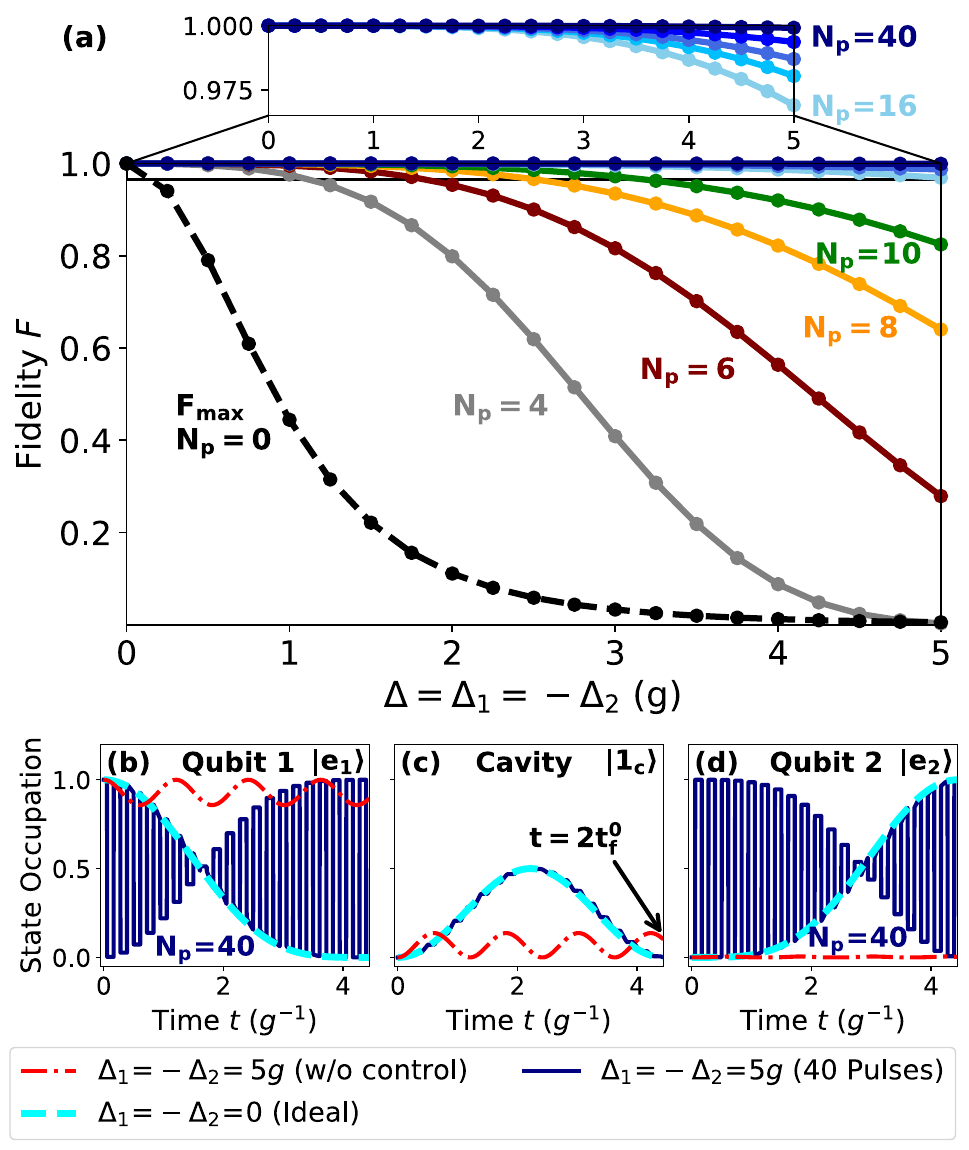}
    \caption{\textbf{(a)}: POCEG fidelity $F$ as a function of the number of pulses for a range of detuning combinations where the qubits are oppositely detuned from the cavity. The dashed black line represents the maximum fidelity achieved without the protocol.  \textbf{(b, c, d)}: Evolution of the occupation of the excited state of the qubits and the first excited state of the cavity as a function of time when the excitation is initially in \textit{Qubit 1}, with the POCEG (navy blue lines) and without it ( dashed red lines). The cyan lines represent the ideal case where both qubits are resonant with the cavity. The corresponding state transfer occurs after a time $t_f^0$ and its plot is rescaled here to match the $2 t^0_f$ timescale of the system under POCEG.
    }
    \label{fig:PP_Main}
\end{figure}

To address this degradation of the fidelity, we introduce POCEG (protocol for optimal cavity-enabled gates). For a cavity with low damping rate, the qubits are simultaneously subjected to a sequence of $\pi_x$ pulses (assumed to be instantaneous) at the cavity frequency $\omega_c$ as per the Carr-Purcell sequence ($\tau/2-\tau-...-\tau-\tau/2$) where $\tau$ is the inter-pulse delay. The coupling $g$ of the qubits with the cavity is  turned off after each odd-numbered pulse to prevent unwanted population in the higher excitation sectors \cite{Beaudoin_2017}. These modulations are schematically illustrated by the gray-solid lines in Fig.~\ref{fig:system} (a). 
We track the dynamics of the system by diagonalizing the piecewise time-independent Hamiltonian exactly and evolving the initial state from time $t=0$ up to some maximum time. For details on this numerical solution, see \ref{sec:Appendix_Numerical_Method}. Furthermore, to gain some insight into the mechanism of the fidelity enhancement, we derive the perturbative solution based on the average Hamiltonian theory.  Under the above-described POCEG sequence, one can obtain the average Hamiltonian of the system in the frame rotating at the cavity frequency:
\begin{eqnarray}
\bar{H} &=& \frac{g}{2}\bigg(a_c^\dagger \sigma^-_1 + a_c \sigma^+_1 + a_c^\dagger \sigma^-_2 + a_c \sigma^+_2\bigg) \nonumber \\
&-& \frac{g \tau^2}{48} \bigg( \Delta^2_1 \bigg(a_c^\dagger \sigma^-_1 + a_c \sigma^+_1 \bigg) + \Delta^2_2 \bigg(a_c^\dagger \sigma^-_2 + a_c \sigma^+_2 \bigg) \bigg) \nonumber \\
&-& \frac{g^2 \tau^2}{24} \bigg(\Delta_1 \bigg(a_c^\dagger a_c + \frac{1}{2} \bigg)\sigma^z_1 + \Delta_2 \bigg(a_c^\dagger a_c + \frac{1}{2} \bigg)\sigma^z_2 \bigg) \nonumber \\
&-& \frac{g^2 \tau^2}{48} \bigg( (\Delta_1 + \Delta_2)(\sigma^+_1 \sigma^-_2 + \sigma^-_1 \sigma^+_2) \bigg)
\label{eq:average_hamiltonian_main}
\end{eqnarray}
One can see that the leading term does not feature the detuning while all first and second order terms in $\Delta_{1,2}$ appear with factors $\sim\tau^2$. 
Thus, to leading order, the qubits are made resonant with the cavity and, as the inter-pulse delay is shortened, the effect of the detuning is further suppressed from the evolution of the system. For the derivation of Eq.~(\ref{eq:average_hamiltonian_main}), see \ref{sec:Appendix_Avg_H}.

\begin{figure}[t] 
    \centering
    \includegraphics[width=\columnwidth]{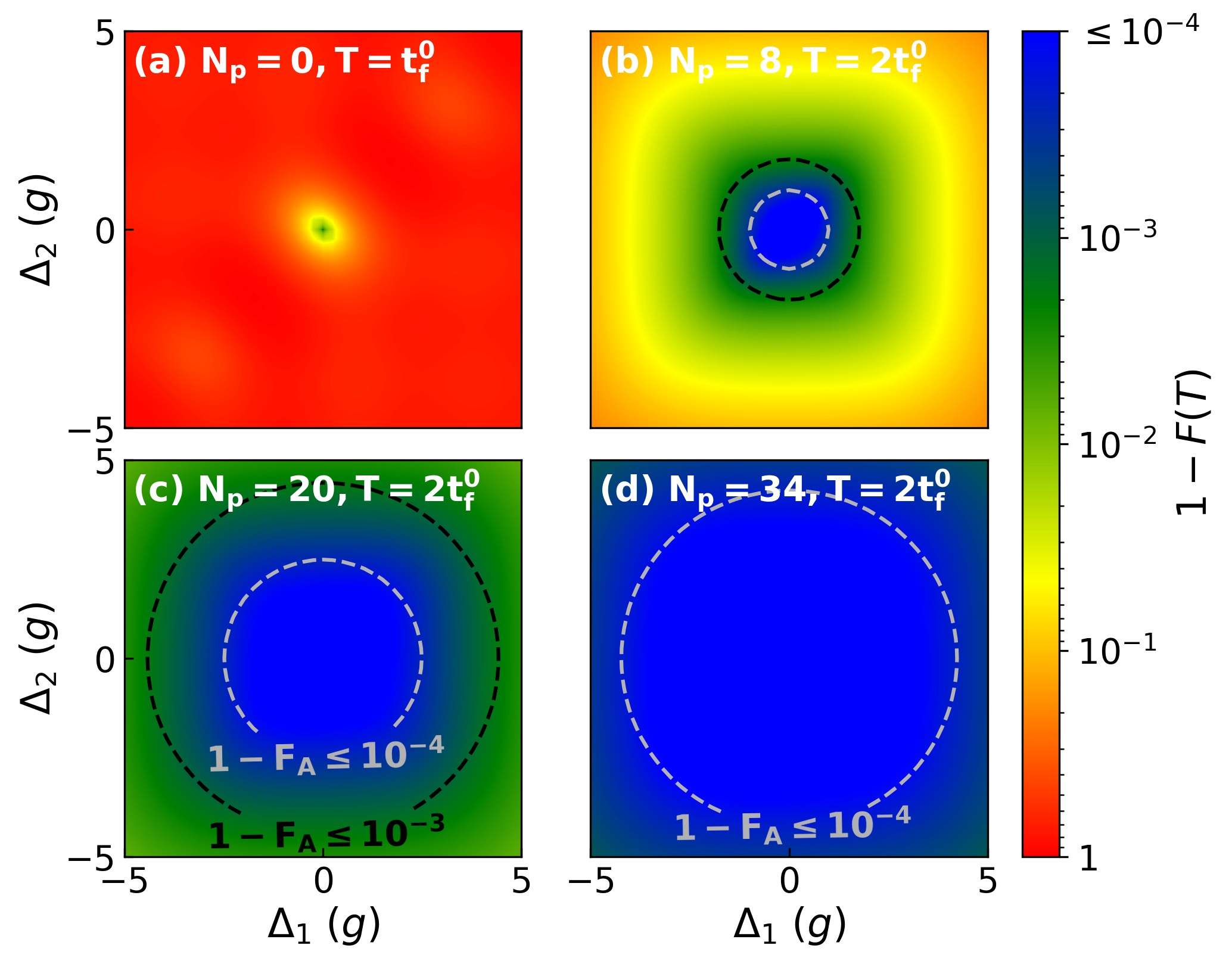}
    \caption{Performance of the POCEG measured as the infidelity $1-F$ as a function of all  detuning combinations ($\Delta_1,\Delta_2$) that represent the quasistatic regime for a noise bandwidth $10g$ centered at the cavity frequency. \textbf{(a):} Without POCEG. \textbf{(b, c, d):} With POCEG for different number of pulses $N_p$ applied periodically within a time $T=2t^0_f$. The dashed black and gray lines represent boundaries within which the analytically derived infidelities are $1-F_A=10^{-3}$ and $1-F_A=10^{-4}$ respectively.}
    \label{fig:Performance_Quasistatic_Detuning}
\end{figure}

\section{Results}
\label{sec:Results}

\noindent Since POCEG combines pulses applied on the qubits with the periodic decoupling of the qubits from the cavity, the minimum evolution time to achieve the desired target state is $2t_f^0$. We thus analyze the fidelity for pairs of detunings $(\Delta_1, \Delta_2)$ as a function of the number of pulses $N_p$ applied between $t=0$ to $t = 2t_f^0$.  

Figure \ref{fig:PP_Main} (a) shows the fidelity $F$ of the state transfer as a function of $\Delta$ with $\Delta_1 = -\Delta_2 = \Delta$, representing oppositely detuned qubits. The dashed line shows the exponential suppression of $F$ with increasing $\Delta$ for the system in the absence of any control. As $N_p$ is gradually increased, we observe a dramatic increase of the $\Delta$ values for which the fidelity remains close to unity. As $N_p$ is increased beyond 16, $F$ remains above 95\% and approaches 99.9\% when $N_p=40$ for all $\Delta$ values between $0$ and $5g$. Figures \ref{fig:PP_Main} (b), (c), and (d) show the time evolution for the occupation of $|e_1\rangle$, $|1_c\rangle$, and $|e_2\rangle$. Where the evolution would be represented by the dashed red lines in the absence of the control protocol, one can readily see (navy blue lines) that the protocol essentially produces for the detuned qubits, here $\Delta_1 = -\Delta_2 = \Delta = 5g$, an evolution that is akin to that of a fully resonant system with no control protocol (cyan lines). Except that while the state transfer for the freely evolving fully resonant system would occur after a time $t_f^0$, the transfer under POCEG occurs at time $2t_f^0$. 

\begin{figure}[t] 
    \centering
    \includegraphics[width=\columnwidth]{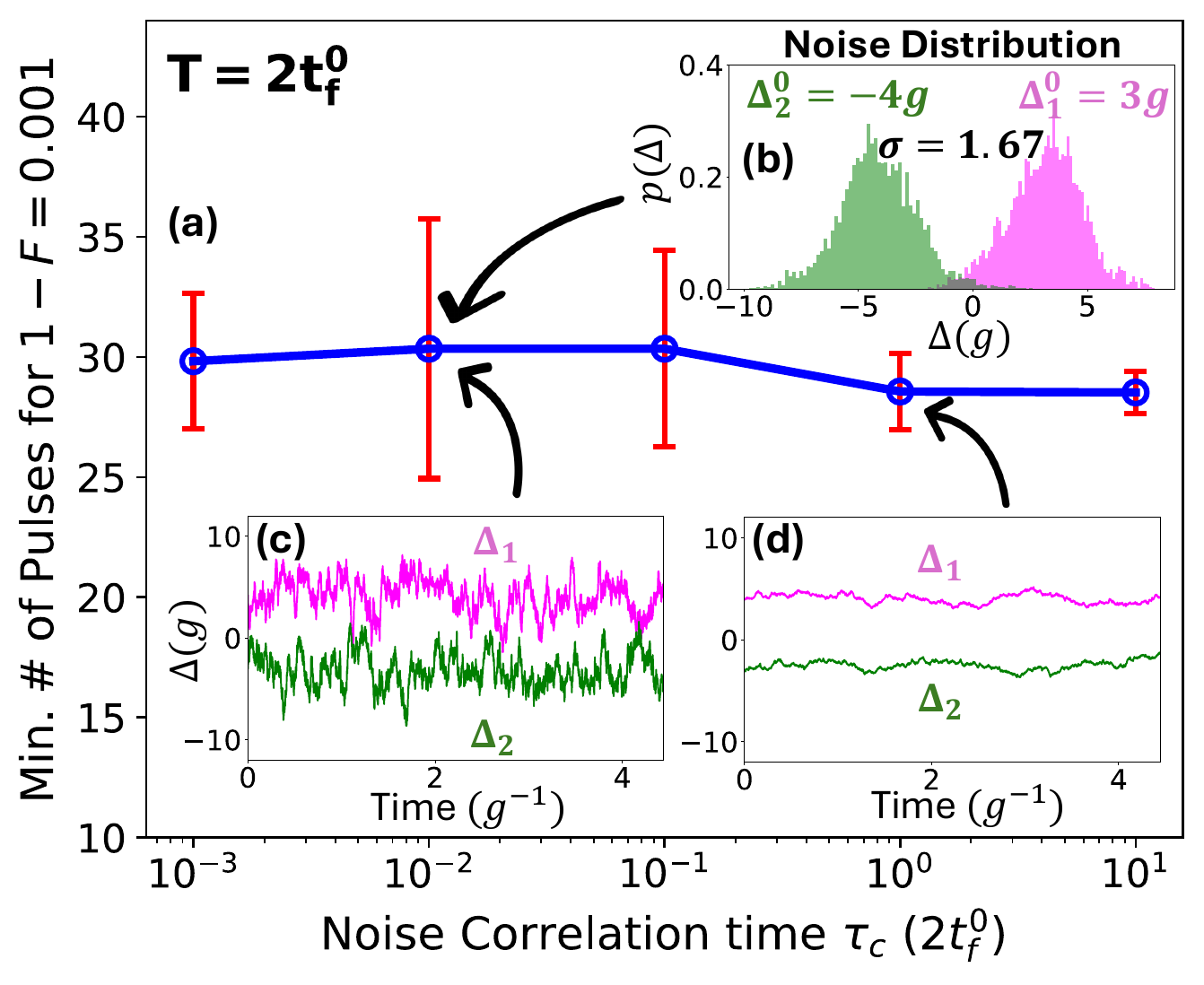}
    \caption{\textbf{(a):} Performance of the POCEG measured as the number of pulses 
    required for $F=0.999$  as a function of the correlation time $\tau_c$ of the noise, ranging from the white noise regime to the quasistatic noise regime. Inset \textbf{(b)} shows a distribution of the detunings $\Delta_1$ and $\Delta_2$ with mean values $\Delta_1^0=3g$ and $\Delta_2^0=-4g$ and standard deviation $\sigma=1.67$, for $\tau_c=0.02t_f^0$, from which its instantaneous detunings are sampled. Insets \textbf{(c)} and \textbf{(d)} show the detunings as a function of time over the protocol transfer time $T=2t_f^0$ for noise correlation times $\tau_c=0.02t_f^0$ and $\tau_c=2t_f^0$ respectively.}
    \label{fig:Noisy_Protocol_Performance}
\end{figure}

\begin{figure}[t] 
    \centering
    \includegraphics[width=\columnwidth]{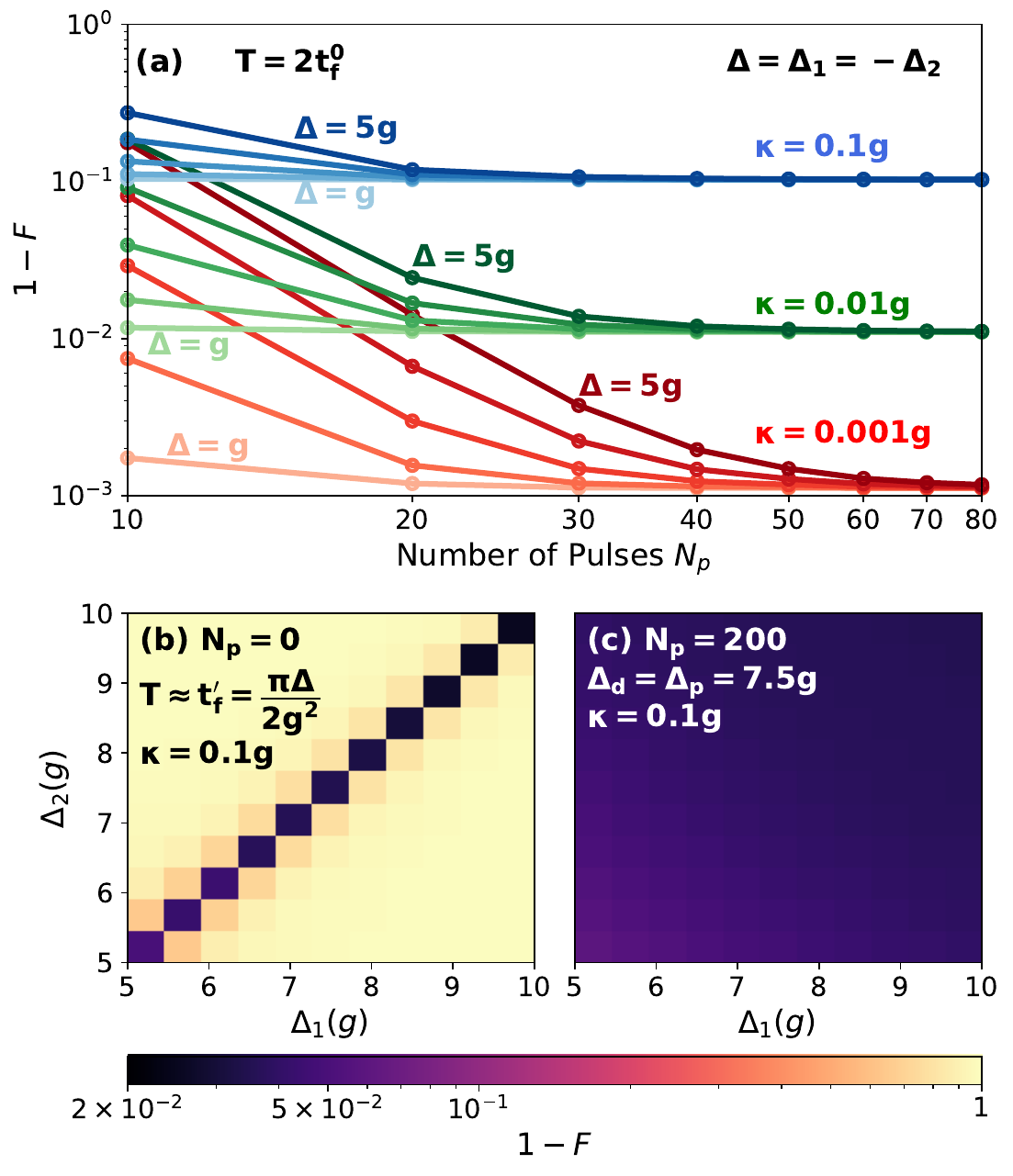}
    \caption{Infidelity $1-F$ as a function of the number of pulses $N_p$ for different values of the cavity damping rate $\kappa$ and detunings $\Delta_1=-\Delta_2=\Delta$ ranging from $g$ to $5g$. \textbf{(a):} Performance of POCEG for different values of $\kappa$. \textbf{(b):} State transfer fidelity as a function of the detunings $\Delta_1$ and $\Delta_2$ in the dispersive regime, for $\kappa=0.1g$, emphasizing that dispersive transfer yields higher fidelity relative to resonant transfer when the cavity damping is large. However, fidelity rapidly drops off for $\Delta_1 \ne \Delta_2$. \textbf{(c):} Performance of POCEG-D for $\kappa=0.1g$ when the qubits are dispersively coupled to the cavity, here at a detuning $\Delta_d=7.5g$, to mitigate cavity loss.}
    \label{fig:Damped_Protocol_Performance}
\end{figure}

To more broadly characterize the effect of the protocol, Fig. \ref{fig:Performance_Quasistatic_Detuning} shows the fidelity for any combination of detunings with $-5g \le \Delta_1, \; \Delta_2 \le 5g$. Panel (a) illustrates the uncontrolled system for which only a small region around $(\Delta_1 = 0, \Delta_2 = 0)$ shows optimal fidelity. Panels (b), (c) and (d) indicate that the region of optimal fidelity gradually increases with the number of pulses. Using a perturbative expansion on the average Hamiltonian (\ref{eq:average_hamiltonian_main}) as outlined in \ref{sec:Appendix_Fidelity}, we obtain the following approximate expression for the fidelity as a function of the interpulse delay:
\begin{equation}
    1 - F_A  = \dfrac{\pi^2 \tau^4}{4608} (\Delta_1^2+\Delta_2^2)^2.
    \label{eq:InfidelityAverageHamiltonian}
\end{equation}
The black and gray dashed lines correspond respectively to the infidelity thresholds of $10^{-3}$ and $10^{-4}$.

The solutions described so far correspond to static detunings between the qubits and the cavity. Next, we consider the case of two qubits with detunings that fluctuate according to an Ornstein-Uhlenbeck process. Specifically, the detunings fluctuate randomly in time following a Gaussian distribution centered at $\Delta_{1}^0$ for \textit{Qubit 1} and at $\Delta_{2}^0$ for \textit{Qubit 2}, both with correlation time $\tau_c$ and standard deviation $\sigma$. In the main panel of Fig.~\ref{fig:Noisy_Protocol_Performance} we show the number of pulses that are necessary to achieve an infidelity less than $10^{-3}$ as a function of the correlation time of the distribution for $\Delta_{1}^0 = 3g$ and $\Delta_{2}^0 = -4g$ with $\sigma = 1.67 $, covering a broadening range that is several times larger than that of recent experiments \cite{DijkemaVandersypen_NatPhys2025, PhysRevX.12.021026}. The figure shows that the protocol performs optimally for correlation times ranging from $10^{-3} T$ to $10 T$, requiring only about 30 pulses to achieve the required fidelity after time $T=2t_f^0$.

We now consider the effect of a cavity with finite damping rate $\kappa$. We examine the minimum infidelity $1 - F$  for a range of detunings as a function of the number of pulses using the master equation \cite{Cohen_Tannoudji_Book1992}. Fig.\ref{fig:Damped_Protocol_Performance} (a) shows that for $\kappa =  10^{-3}g$, the infidelity saturates at $10^{-3}$ requiring a number of pulses that depends on the values of the detunings. The figure considers the case of oppositely detuned qubits, $\Delta_1 = - \Delta_2 = \Delta$, and the maximum number of pulses required to achieve $10^{-3}$ infidelity is $\sim 70$. Generally, the finite damping rate introduces a threshold in the achievable infidelity. This threshold increases with the damping strength and is achieved for fewer pulses as $\kappa$ increases. For $\kappa = 0.1 g$ \cite{DijkemaVandersypen_NatPhys2025, PhysRevX.12.021026}, $1 - F$ saturates for the same range of $\Delta$ values at $\sim 10^{-1}$. This is expected as the protocol with pulses at the cavity frequency aims to make the qubits resonant with the cavity and thus will be susceptible to the deleterious effects of the photon loss. As is typical for such cavities, operating in the dispersive regime is preferable. However, without spectral enhancement, optimal performance is only achieved when both qubits, although far-detuned from the cavity, are at the same frequency, i.e $\Delta_1 = \Delta_2 = \Delta > g$. This is pictured in panel (b) of Fig.\ref{fig:Damped_Protocol_Performance} where the maximum fidelity obtained after time $t_f'=\pi \Delta/2g^2$ is observed along the diagonal of the $(\Delta_1, \Delta_2)$ plane.  To enhance the fidelity of the operation for such a cavity, we propose a variation of our protocol, POCEG-D, which consists in applying pulses at a frequency that is far-detuned from the cavity. Since this protocol now makes both qubits identical and while keeping them far-detuned from the cavity, this variation of the protocol does not require modulating the coupling to the cavity. However, similar to its resonant variation, it optimizes the operation for broad combinations  $(\Delta_1, \Delta_2)$ of detunings among the emitters, as pictured in panel (c) of Fig. \ref{fig:Damped_Protocol_Performance}. Applying pulses to the qubits at detuning $\Delta_p=\Delta_d$ enables effective dispersive transfer even when $\Delta_1 \ne \Delta_2$, yielding a higher fidelity ($\approx 94 \%$ to $97 \%$) for $\kappa=0.1g$ compared to its counterpart in panel (a) ($\approx 90 \% $). The time for state transfer here is $2t_f' \approx 11.3 t_f^0$. As seen in panel (b), operating POCEG-D at a higher $\Delta_d$ will further increase the achievable fidelity.\\

\section{Conclusion}
\label{sec:Conclusion}
\noindent To address the challenging scalability issue of photon-mediated qubit two-qubit gates in QIP, we have introduced a protocol dubbed POCEG (protocol for optimal cavity-enabled gates) that is shown, through a combination of analytical and numerical solutions, to overcome spectral differences between quantum bits and to dramatically enhance the fidelity of the cavity-mediated quantum state transfer operation between disparate quantum emitters. For a cavity with low damping rate, we apply a sequence of pulses to the qubits at the frequency of the cavity while periodically modulating the coupling of the qubits to the cavity. Alternatively, in the case of a cavity with a large damping rate, we operate the system in the dispersive regime and address spectral differences between the qubits by applying the pulses at a frequency far-detuned from the cavity. In both cases, we find for the quantum state transfer between the two qubits that, with a modest inter-pulse delay, the fidelity that would otherwise be strongly suppressed by the spectral mismatch of the qubits can be increased beyond 99.9\%. These protocols represent a promising avenue to bring two-qubit gates between solid state systems across their scalability threshold.

\section*{Acknowledgment} 
\noindent We acknowledge support from the National Science Foundation under Grants No. PHY-2014023 and No. QIS-2328752. This work was also partially supported by the Microelectronics Commons Program, a DoW initiative, under award number N00164-23-9-G061 and AFRL/RI under contract number SA10032026051364. This document is approved for
public release, distribution unlimited. PA 26-T-2281. We thank V. V. Dobrovitski and X. Hu for useful discussions.

\bibliography{References}

\onecolumngrid
\appendix
\newpage

\clearpage
\section*{Supplemental Material}
\addcontentsline{toc}{section}{Supplemental Material}

\appendix
\renewcommand{\thesection}{SM-\Roman{section}}
\renewcommand{\theequation}{\thesection-\arabic{equation}}
\renewcommand{\appendixname}{}

\section{Average Hamiltonian of the system under POCEG}
\label{sec:Appendix_Avg_H}

\noindent For a time-dependent Hamiltonian where the Hamiltonian does not commute across time, the evolution operator involves a time-ordered integral. In such cases, the average Hamiltonian theory converts the time-dependent Hamiltonian into an effective time-independent Hamiltonian over a defined time interval \cite{Avg_H_Viola_1999_PRL, brinkmann_avg_Hamiltonian_2016}.
\begin{eqnarray}
    U(t) = \mathcal{T}\mathrm{exp}\bigg(-\mathrm{i} \int^0_t dt^{'} H(t^{'}) \bigg) \approx \mathrm{exp} \bigg( -\mathrm{i}\bar{H}t \bigg)
    \label{eq:Avg_H_1}
\end{eqnarray}

\noindent The average Hamiltonian $\bar{H}$ is given by the Magnus expansion \cite{magnus1954exponential},
\begin{eqnarray}
\bar{H} &=& \bar{H}^{(0)} + \bar{H}^{(1)} +\bar{H}^{(2)} + ... 
\label{eq:Avg_H_2}
\end{eqnarray}

\noindent With,
\begin{eqnarray}
\bar{H}^{(0)} &=& \frac{1}{2\tau} \int_0^{2\tau} H(t_1) \, dt_1 \nonumber \\
\bar{H}^{(1)} &=& \frac{1}{4\mathrm{i}\tau} \int_0^{2\tau} dt_1 \int_0^{t_1} dt_2 \, [H(t_1), H(t_2)] \nonumber\\
\bar{H}^{(2)} &=& -\frac{1}{12\tau} \int_0^{2\tau} dt_1 \int_0^{t_1} dt_2 \int_0^{t_2} dt_3 \, \{[H(t_1), [H(t_2), H(t_3)]] + [[H(t_1), H(t_2)], H(t_3), ] \}
\label{eq:Avg_H_3}
\end{eqnarray}

\noindent The POCEG Hamiltonian (\ref{eq:Hamiltonian_POCEG}) in the rotating wave approximation (RWA) and in the rotating frame of the cavity is:
\begin{eqnarray}
   H = \frac{\Delta_1}{2}\sigma^z_1 + \frac{\Delta_2}{2}\sigma^z_2 + g(t) (a_c^\dagger \sigma^-_1 + a_c \sigma^+_1 + a_c^\dagger \sigma^-_2 + a_c \sigma^+_2) + \Omega(t)(\sigma^x_1 + \sigma^x_2)
\label{eq:Hamiltonian_POCEG_RWA} 
\end{eqnarray}

\noindent Moving to the toggling frame,
\begin{eqnarray}
\tilde{H} &=& U^\dagger_c(t)H(t)U_c(t) \nonumber \\
&=& \mathrm{exp}\bigg(\mathrm{i}\frac{\Omega(t)t}{2}(\sigma^x_1 + \sigma^x_2) \bigg) \bigg(\frac{\Delta_1}{2}\sigma^z_1 + \frac{\Delta_2}{2}\sigma^z_2 + g(t) (a_c^\dagger \sigma^-_1 + a_c \sigma^+_1 + a_c^\dagger \sigma^-_2 + a_c \sigma^+_2) \bigg) \nonumber \\ 
&\quad& \times\mathrm{exp}\bigg(-\mathrm{i}\frac{\Omega(t)t}{2}(\sigma^x_1 + \sigma^x_2) \bigg)
\label{eq:Avg_H_4}
\end{eqnarray}

\noindent Noting that for an instantaneous $\pi$-pulse, $\Omega t = \pi$, we have,
\begin{eqnarray}
\exp(\mathrm{i} \frac{\pi}{2} \sigma_x)\sigma_z \exp(-\mathrm{i} \frac{\pi}{2} \sigma_x) &=& -\sigma_z \nonumber \\
\exp(\mathrm{i} \frac{\pi}{2} \sigma_x)\sigma_+ \exp(-\mathrm{i} \frac{\pi}{2} \sigma_x) &=& \sigma_- \nonumber \\
\exp(\mathrm{i} \frac{\pi}{2} \sigma_x )\sigma_- \exp(-\mathrm{i} \frac{\pi}{2} \sigma_x) &=& \sigma_+
\label{eq:Avg_H_5}
\end{eqnarray}

\noindent Using the above relations, the Hamiltonian in the toggling frame becomes:
\begin{eqnarray}
\tilde{H}_{\mathrm{even}} &=& \frac{\Delta_1}{2}\sigma^z_1 + \frac{\Delta_2}{2}\sigma^z_2 + g(t)(a_c^\dagger \sigma^-_1 + a_c \sigma^+_1 + a_c^\dagger \sigma^-_2 + a_c \sigma^+_2) \quad\quad n(t) \quad \mathrm{even}, \nonumber \\
\tilde{H}_{\mathrm{odd}} &=& -\frac{\Delta_1}{2}\sigma^z_1 - \frac{\Delta_2}{2}\sigma^z_2 + g(t)(a_c^\dagger \sigma^-_1 + a_c \sigma^+_1 + a_c^\dagger \sigma^-_2 + a_c \sigma^+_2) \quad\quad n(t) \quad \mathrm{odd}
\label{eq:Avg_H_6}
\end{eqnarray}

\noindent where $n(t)$ is the number of $\pi$-pulses applied before time $t$. In $\tilde{H}_{\mathrm{odd}}$, the interaction term is not excitation-conserving. To address this problem, we turn off the cavity after an odd number of pulses. This gives the final form of our engineered Hamiltonian,
\begin{eqnarray}
\tilde{H}_{\mathrm{even}} &=& \frac{\Delta_1}{2}\sigma^z_1 + \frac{\Delta_2}{2}\sigma^z_2 + (a_c^\dagger \sigma^-_1 + a_c \sigma^+_1 + a_c^\dagger \sigma^-_2 + a_c \sigma^+_2) \quad\quad n(t) \quad \mathrm{even}, \nonumber \\
\tilde{H}_{\mathrm{odd}} &=& -\frac{\Delta_1}{2}\sigma^z_1 - \frac{\Delta_2}{2}\sigma^z_2 \quad\quad n(t) \quad \mathrm{odd}
\label{eq:Avg_H_7}
\end{eqnarray}

\noindent We calculate the average Hamiltonian up to the first correction using the engineered Hamiltonian from (\ref{eq:Avg_H_7}) over a $2\tau$ interval.
\begin{eqnarray}
\bar{H}^{(0)} &=& \frac{1}{2\tau} \bigg(\int_0^{\frac{\tau}{2}} \tilde{H_1} \, dt_1 + \int_{\frac{\tau}{2}}^{\frac{3\tau}{2}} \tilde{H_2} \, dt_1 + \int_{\frac{3\tau}{2}}^{2\tau} \tilde{H_1} \, dt_1\bigg) \nonumber \\
&=& \frac{1}{2\tau} \bigg(\frac{\tau}{2} \bigg(\frac{\Delta_1}{2}\sigma^z_1 + \frac{\Delta_2}{2}\sigma^z_2 + g(a_c^\dagger \sigma^-_1 + a_c \sigma^+_1 + a_c^\dagger \sigma^-_2 + a_c \sigma^+_2 \bigg) \bigg) \nonumber \\
&\quad\quad\quad\quad& -\tau \bigg(\frac{\Delta_1}{2}\sigma^z_1 + \frac{\Delta_2}{2}\sigma^z_2 \bigg) \nonumber \\
&\quad\quad\quad\quad& + \frac{\tau}{2} \bigg(\frac{\Delta_1}{2}\sigma^z_1 + \frac{\Delta_2}{2}\sigma^z_2 + g\bigg(a_c^\dagger \sigma^-_1 + a_c \sigma^+_1 + a_c^\dagger \sigma^-_2 + a_c \sigma^+_2\bigg)\bigg)\bigg) \nonumber \\
&=& \frac{g}{2}\bigg(a_c^\dagger \sigma^-_1 + a_c \sigma^+_1 + a_c^\dagger \sigma^-_2 + a_c \sigma^+_2\bigg) \nonumber \\
\label{eq:Avg_H_8}
\end{eqnarray}

\noindent The Hamiltonian in the toggling-frame is given by (\ref{eq:Avg_H_6}) has the following property: $\tilde{H}(t ) = \tilde{H}(2\tau - t )$, corresponding to a symmetric cycle with period $T = 2\tau$. For such symmetric cycles, all odd orders vanish in the Magnus expansion \cite{mansfield1971symmetrized}.
\begin{eqnarray}
\bar{H}^{(1)} = 0
\label{eq:Avg_H_9}
\end{eqnarray}

\noindent The first correction is given by,
\begin{eqnarray}
\bar{H}^{(2)} &=& -\frac{1}{12\tau} \bigg( \int_0^{\frac{\tau}{2}} dt_1 \int_0^{t_1} dt_2 \int_0^{t_2} dt_3 \{[\tilde{H}_{\mathrm{even}}, [\tilde{H}_{\mathrm{even}}, \tilde{H}_{\mathrm{even}}]] + [[\tilde{H}_{\mathrm{even}}, \tilde{H}_{\mathrm{even}}], \tilde{H}_{\mathrm{even}}]\} \nonumber\\
&\quad\quad& + \int_{\frac{\tau}{2}}^{\frac{3\tau}{2}} dt_1 \int_0^{\frac{\tau}{2}} dt_2 \int_0^{t_2} dt_3 \{[\tilde{H}_{\mathrm{odd}}, [\tilde{H}_{\mathrm{even}}, \tilde{H}_{\mathrm{even}}]] + [[\tilde{H}_{\mathrm{odd}}, \tilde{H}_{\mathrm{even}}], \tilde{H}_{\mathrm{even}}]\} \nonumber \\
&\quad\quad& + \int_{\frac{\tau}{2}}^{\frac{3\tau}{2}} dt_1 \int_{\frac{\tau}{2}}^{t_1} dt_2 \int_0^{\frac{\tau}{2}} dt_3 \{[\tilde{H}_{\mathrm{odd}}, [\tilde{H}_{\mathrm{odd}}, \tilde{H}_{\mathrm{even}}]] + [[\tilde{H}_{\mathrm{odd}}, \tilde{H}_{\mathrm{odd}}], \tilde{H}_{\mathrm{even}}]\} \nonumber \\
&\quad\quad& + \int_{\frac{\tau}{2}}^{\frac{3\tau}{2}} dt_1 \int_{\frac{\tau}{2}}^{t_1} dt_2 \int_{\frac{\tau}{2}}^{t_2} dt_3 \{[\tilde{H}_{\mathrm{odd}}, [\tilde{H}_{\mathrm{odd}}, \tilde{H}_{\mathrm{odd}}]] + [[\tilde{H}_{\mathrm{odd}}, \tilde{H}_{\mathrm{odd}}], \tilde{H}_{\mathrm{odd}}]\} \nonumber \\
&\quad\quad& + \int_{\frac{3\tau}{2}}^{2\tau} dt_1 \int_{0}^{\frac{\tau}{2}} dt_2 \int_{0}^{t_2} dt_3 \{[\tilde{H}_{\mathrm{even}}, [\tilde{H}_{\mathrm{even}}, \tilde{H}_{\mathrm{even}}]] + [[\tilde{H}_{\mathrm{even}}, \tilde{H}_{\mathrm{even}}], \tilde{H}_{\mathrm{even}}]\} \nonumber \\
&\quad\quad& + \int_{\frac{3\tau}{2}}^{2\tau} dt_1 \int_{\frac{\tau}{2}}^{\frac{3\tau}{2}} dt_2 \int_{0}^{\frac{\tau}{2}} dt_3 \{[\tilde{H}_{\mathrm{even}}, [\tilde{H}_{\mathrm{odd}}, \tilde{H}_{\mathrm{even}}]] + [[\tilde{H}_{\mathrm{even}}, \tilde{H}_{\mathrm{odd}}], \tilde{H}_{\mathrm{even}}]\} \nonumber \\
&\quad\quad& + \int_{\frac{3\tau}{2}}^{2\tau} dt_1 \int_{\frac{\tau}{2}}^{\frac{3\tau}{2}} dt_2 \int_{\frac{\tau}{2}}^{t_2} dt_3 \{[\tilde{H}_{\mathrm{even}}, [\tilde{H}_{\mathrm{odd}}, \tilde{H}_{\mathrm{odd}}]] + [[\tilde{H}_{\mathrm{even}}, \tilde{H}_{\mathrm{odd}}], \tilde{H}_{\mathrm{odd}}]\} \nonumber \\
&\quad\quad& + \int_{\frac{3\tau}{2}}^{2\tau} dt_1 \int_{\frac{3\tau}{2}}^{t_1} dt_2 \int_{0}^{\frac{\tau}{2}} dt_3 \{[\tilde{H}_{\mathrm{even}}, [\tilde{H}_{\mathrm{even}}, \tilde{H}_{\mathrm{even}}]] + [[\tilde{H}_{\mathrm{even}}, \tilde{H}_{\mathrm{even}}], \tilde{H}_{\mathrm{even}}]\} \nonumber \\
&\quad\quad& + \int_{\frac{3\tau}{2}}^{2\tau} dt_1 \int_{\frac{3\tau}{2}}^{t_1} dt_2 \int_{\frac{\tau}{2}}^{\frac{3\tau}{2}} dt_3 \{[\tilde{H}_{\mathrm{even}}, [\tilde{H}_{\mathrm{even}}, \tilde{H}_{\mathrm{odd}}]] + [[\tilde{H}_{\mathrm{even}}, \tilde{H}_{\mathrm{even}}], \tilde{H}_{\mathrm{odd}}]\} \nonumber \\
&\quad\quad& + \int_{\frac{3\tau}{2}}^{2\tau} dt_1 \int_{\frac{3\tau}{2}}^{t_1} dt_2 \int_{\frac{3\tau}{2}}^{t_2} dt_3 \{[\tilde{H}_{\mathrm{even}}, [\tilde{H}_{\mathrm{even}}, \tilde{H}_{\mathrm{even}}]] + [[\tilde{H}_{\mathrm{even}}, \tilde{H}_{\mathrm{even}}], \tilde{H}_{\mathrm{even}}]\}\bigg)
\label{eq:Avg_H_10}
\end{eqnarray}

\noindent Noting that the integrands are now time-independent, they can be evaluated separately from the commutator terms. It is easy to see which commutators drop out. The simplified expression is as follows.
\begin{eqnarray}
\bar{H}^{(2)} &=& -\frac{1}{12\tau} \bigg(
\frac{\tau^3}{8} [[\tilde{H}_{\mathrm{odd}}, \tilde{H}_{\mathrm{even}}], \tilde{H}_{\mathrm{even}}] \nonumber \\
&\quad\quad& + \frac{\tau^3}{4} [\tilde{H}_{\mathrm{odd}}, [\tilde{H}_{\mathrm{odd}}, \tilde{H}_{\mathrm{even}}]] \nonumber \\
&\quad\quad& + \frac{\tau^3}{4} \{[\tilde{H}_{\mathrm{even}}, [\tilde{H}_{\mathrm{odd}}, \tilde{H}_{\mathrm{even}}]] + [[\tilde{H}_{\mathrm{even}}, \tilde{H}_{\mathrm{odd}}], \tilde{H}_{\mathrm{even}}]\} \nonumber \\
&\quad\quad& + \frac{\tau^3}{4} [[\tilde{H}_{\mathrm{even}}, \tilde{H}_{\mathrm{odd}}], \tilde{H}_{\mathrm{odd}}] \nonumber \\
&\quad\quad& + \frac{\tau^3}{8} [\tilde{H}_{\mathrm{even}}, [\tilde{H}_{\mathrm{even}}, \tilde{H}_{\mathrm{odd}}]]
\bigg)
\label{eq:Avg_H_11}
\end{eqnarray}

\noindent Using $[A, B] = -[B, A]$, the above expression can be reduced to,
\begin{eqnarray}
\bar{H}^{(2)} &=& -\frac{1}{12\tau} \bigg(
\bigg(\frac{\tau^3}{8} - \frac{\tau^3}{4} - \frac{\tau^3}{4} + \frac{\tau^3}{8} \bigg) [[\tilde{H}_{\mathrm{odd}}, \tilde{H}_{\mathrm{even}}], \tilde{H}_{\mathrm{even}}] + \bigg(\frac{\tau^3}{4} + \frac{\tau^3}{4} \bigg) [\tilde{H}_{\mathrm{odd}}, [\tilde{H}_{\mathrm{odd}}, \tilde{H}_{\mathrm{even}}]] \bigg) \nonumber \\
&=& \frac{\tau^2}{48} \bigg[\bigg[\tilde{H}_{\mathrm{odd}}, \tilde{H}_{\mathrm{even}}\bigg], \tilde{H}_{\mathrm{even}}\bigg]
- \frac{\tau^2}{24} \bigg[\tilde{H}_{\mathrm{odd}}, \bigg[\tilde{H}_{\mathrm{odd}}, \tilde{H}_{\mathrm{even}}\bigg]\bigg]
\label{eq:Avg_H_12}
\end{eqnarray}

\noindent Next, we evaluate the commutator terms using the following identities.
\begin{eqnarray}
&[\sigma_-, \sigma_z] = 2\sigma_- \nonumber \\
&[\sigma_+, \sigma_z] = -2\sigma_+ \nonumber \\
&[\sigma_+, \sigma_- ] = \sigma_z \nonumber \\
&[a, a^{\dagger} ] = 1\nonumber \\
&[A, B] = -[B, A]
\label{eq:Avg_H_13}
\end{eqnarray}

\begin{eqnarray}
\bigg[\tilde{H}_{\mathrm{odd}}, \tilde{H}_{\mathrm{even}}\bigg] &=& \bigg[\bigg(-\frac{\Delta_1}{2}\sigma^z_1 - \frac{\Delta_2}{2}\sigma^z_2 \bigg), \bigg(\frac{\Delta_1}{2}\sigma^z_1 + \frac{\Delta_2}{2}\sigma^z_2 + g(a_c^\dagger \sigma^-_1 + a_c \sigma^+_1 + a_c^\dagger \sigma^-_2 + a_c \sigma^+_2) \bigg) \bigg] \nonumber \\
&=& -\frac{\Delta_1 g}{2}\bigg[\sigma^z_1, a_c^\dagger \sigma^-_1 \bigg] - \frac{\Delta_1 g}{2}\bigg[\sigma^z_1, a_c \sigma^+_1 \bigg] -\frac{\Delta_2 g}{2}\bigg[\sigma^z_2, a_c^\dagger \sigma^-_2 \bigg] - \frac{\Delta_2 g}{2}\bigg[\sigma^z_2, a_c \sigma^+_2 \bigg] \nonumber \\
&=& -\frac{\Delta_1 g}{2}\bigg(a_c^\dagger \bigg(-2\sigma^-_1\bigg) +  a_c \bigg(2\sigma^+_1\bigg)\bigg) 
-\frac{\Delta_2 g}{2}\bigg(a_c^\dagger \bigg(-2\sigma^-_2\bigg) +  a_c \bigg(2\sigma^+_2\bigg)\bigg) \nonumber \\
&=& \Delta_1 g\bigg(a_c^\dagger \sigma^-_1 - a_c \sigma^+_1 \bigg) + \Delta_2 g\bigg(a_c^\dagger \sigma^-_2 - a_c \sigma^+_2 \bigg)
\label{eq:Avg_H_14}
\end{eqnarray}

\begin{eqnarray}
\bigg[\tilde{H}_{\mathrm{odd}}, \bigg[\tilde{H}_{\mathrm{odd}}, \tilde{H}_{\mathrm{even}}\bigg]\bigg] &=& \bigg[\bigg(-\frac{\Delta_1}{2}\sigma^z_1 - \frac{\Delta_2}{2}\sigma^z_2 \bigg), \bigg( \Delta_1 g\bigg(a_c^\dagger \sigma^-_1 - a_c \sigma^+_1 \bigg) + \Delta_2 g\bigg(a_c^\dagger \sigma^-_2 - a_c \sigma^+_2 \bigg) \bigg) \bigg] \nonumber \\
&=& -\frac{\Delta^2_1 g}{2} \bigg( \bigg[\sigma^z_1, a_c^\dagger \sigma^-_1 \bigg] - \bigg[\sigma^z_1, a_c \sigma^+_1 \bigg] \bigg) -\frac{\Delta^2_2 g}{2} \bigg( \bigg[\sigma^z_2, a_c^\dagger \sigma^-_2 \bigg] - \bigg[\sigma^z_2, a_c \sigma^+_2 \bigg] \bigg) \nonumber \\
&=& -\frac{\Delta^2_1 g}{2} \bigg(a_c^\dagger (-2\sigma^-_1) - a_c(2\sigma^+_1)\bigg) -\frac{\Delta^2_2 g}{2} \bigg(a_c^\dagger (-2\sigma^-_2) - a_c(2\sigma^+_2)\bigg) \nonumber \\
&=& \Delta^2_1 g \bigg(a_c^\dagger \sigma^-_1 + a_c \sigma^+_1 \bigg) + \Delta^2_2 g \bigg(a_c^\dagger \sigma^-_2 + a_c \sigma^+_2 \bigg)
\label{eq:Avg_H_15}
\end{eqnarray}

\begin{eqnarray}
\bigg[\bigg[\tilde{H}_{\mathrm{odd}}, \tilde{H}_{\mathrm{even}}\bigg], \tilde{H}_{\mathrm{even}} \bigg]
&=& \bigg[ \bigg( \Delta_1 g\bigg(a_c^\dagger \sigma^-_1 - a_c \sigma^+_1 \bigg) + \Delta_2 g\bigg(a_c^\dagger \sigma^-_2 - a_c \sigma^+_2 \bigg) \bigg), \nonumber \\
&\quad\quad& \bigg(\frac{\Delta_1}{2}\sigma^z_1 + \frac{\Delta_2}{2}\sigma^z_2 + g(a_c^\dagger \sigma^-_1 + a_c \sigma^+_1 + a_c^\dagger \sigma^-_2 + a_c \sigma^+_2) \bigg) \bigg] \nonumber \\
&=& \frac{\Delta^2_1 g}{2} \bigg[a_c^\dagger \sigma^-_1, \sigma^z_1 \bigg] + \Delta_1 g^2 \bigg[a_c^\dagger \sigma^-_1, a_c\sigma^+_1 \bigg] + \Delta_1 g^2 \bigg[a_c^\dagger \sigma^-_1, a_c\sigma^+_2 \bigg] \nonumber \\
&\quad\quad& - \frac{\Delta^2_1 g}{2} \bigg[a_c \sigma^+_1, \sigma^z_1 \bigg] - \Delta_1 g^2 \bigg[a_c \sigma^+_1, a_c^\dagger \sigma^-_1 \bigg] - \Delta_1 g^2 \bigg[a_c \sigma^+_1, a^\dagger \sigma^-_2 \bigg] \nonumber \\
&\quad\quad& + \frac{\Delta^2_2 g}{2} \bigg[a_c^\dagger \sigma^-_2, \sigma^z_2 \bigg] + \Delta_2 g^2 \bigg[a_c^\dagger \sigma^-_2, a_c \sigma^+_1 \bigg] + \Delta_2 g^2 \bigg[a_c^\dagger \sigma^-_2, a_c \sigma^+_2 \bigg] \nonumber \\
&\quad\quad& - \frac{\Delta^2_2 g}{2} \bigg[a_c \sigma^+_2, \sigma^z_2 \bigg] - \Delta_2 g^2 \bigg[a_c \sigma^+_2, a_c^\dagger \sigma^-_1 \bigg] - \Delta_2 g^2 \bigg[a \sigma^+_2, a_c^\dagger \sigma^-_2 \bigg] \nonumber \\
&=& \Delta^2_1 g \bigg(a_c^\dagger \sigma^-_1 + a_c \sigma^+_1 \bigg) - 2\Delta_1 g^2 \bigg(a_c^\dagger a_c + \frac{1}{2} \bigg) \sigma^z_1 \nonumber \\ 
&\quad\quad& + \Delta^2_2 g \bigg(a_c^\dagger \sigma^-_2 + a_c \sigma^+_2 \bigg) - 2\Delta_2 g^2 \bigg(a_c^\dagger a_c + \frac{1}{2} \bigg) \sigma^z_2 \nonumber \\
&\quad\quad& - g^2 (\Delta_1 + \Delta_2)(\sigma^+_1 \sigma^-_2 + \sigma^-_1 \sigma^+_2)
\label{eq:Avg_H_16}
\end{eqnarray}

\noindent Plugging (\ref{eq:Avg_H_15}) and (\ref{eq:Avg_H_16}) into (\ref{eq:Avg_H_12}),
\begin{eqnarray}
\bar{H}^{(2)} &=& \frac{\tau^2}{48} \bigg( \Delta^2_1 g \bigg(a_c^\dagger \sigma^-_1 + a_c \sigma^+_1 \bigg) - 2\Delta_1 g^2 \bigg(a_c^\dagger a_c + \frac{1}{2} \bigg) \sigma^z_1 \nonumber \\ 
&\quad\quad& + \Delta^2_2 g \bigg(a_c^\dagger \sigma^-_2 + a_c \sigma^+_2 \bigg) - 2\Delta_2 g^2 \bigg(a_c^\dagger a_c + \frac{1}{2} \bigg) \sigma^z_2 \nonumber \\
&\quad\quad& - g^2 (\Delta_1 + \Delta_2)(\sigma^+_1 \sigma^-_2 + \sigma^-_1 \sigma^+_2)\bigg) \nonumber \\
&\quad\quad& - \frac{\tau^2}{24} \bigg( \Delta^2_1 g \bigg(a_c^\dagger \sigma^-_1 + a_c \sigma^+_1 \bigg) + \Delta^2_2 g \bigg(a_c^\dagger \sigma^-_2 + a_c \sigma^+_2 \bigg) \bigg) \nonumber\\
&=& -\frac{g \tau^2}{48} \bigg( \Delta^2_1 \bigg(a_c^\dagger \sigma^-_1 + a_c \sigma^+_1 \bigg) + \Delta^2_2 \bigg(a_c^\dagger \sigma^-_2 + a_c \sigma^+_2 \bigg) \bigg) \nonumber \\
&\quad\quad& -\frac{g^2 \tau^2}{24} \bigg(\Delta_1 \bigg(a_c^\dagger a_c + \frac{1}{2} \bigg)\sigma^z_1 + \Delta_2 \bigg(a_c^\dagger a_c + \frac{1}{2} \bigg)\sigma^z_2 \bigg) \nonumber \\
&\quad\quad& -\frac{g^2 \tau^2}{48} \bigg( (\Delta_1 + \Delta_2)(\sigma^+_1 \sigma^-_2 + \sigma^-_1 \sigma^+_2) \bigg)
\label{eq:Avg_H_17}
\end{eqnarray}

\noindent Plugging  (\ref{eq:Avg_H_8}), (\ref{eq:Avg_H_9}) and (\ref{eq:Avg_H_17}) into (\ref{eq:Avg_H_2}), and truncating $\bar{H}$ to second order gives the effective time-independent average Hamiltonian for the system.
\begin{eqnarray}
\bar{H} &=& \frac{g}{2}\bigg(a_c^\dagger \sigma^-_1 + a_c \sigma^+_1 + a_c^\dagger \sigma^-_2 + a_c \sigma^+_2\bigg) \nonumber \\
&\quad\quad& -\frac{g \tau^2}{48} \bigg( \Delta^2_1 \bigg(a_c^\dagger \sigma^-_1 + a_c \sigma^+_1 \bigg) + \Delta^2_2 \bigg(a_c^\dagger \sigma^-_2 + a_c \sigma^+_2 \bigg) \bigg) \nonumber \\
&\quad\quad& -\frac{g^2 \tau^2}{24} \bigg(\Delta_1 \bigg(a_c^\dagger a_c + \frac{1}{2} \bigg)\sigma^z_1 + \Delta_2 \bigg(a_c^\dagger a_c + \frac{1}{2} \bigg)\sigma^z_2 \bigg) \nonumber \\
&\quad\quad& -\frac{g^2 \tau^2}{48} \bigg( (\Delta_1 + \Delta_2)(\sigma^+_1 \sigma^-_2 + \sigma^-_1 \sigma^+_2) \bigg)
\label{eq:average_hamiltonian}
\end{eqnarray}

\section{Analytical estimate of the state transfer fidelity under the POCEG protocol}
\label{sec:Appendix_Fidelity}

\noindent An analytical expression for the state transfer fidelity produced by the protocol can be obtained through a perturbative treatment of the average Hamiltonian in (\ref{eq:average_hamiltonian_main}) by noting that the inter-pulse delay $\tau$ is small relative to the other timescales in the system. Terms with the prefactor $\tau^2$ in the average Hamiltonian are treated as a perturbation to the first term which represents perfect coupling. We then have,

\begin{eqnarray}
    H_0 = \frac{g}{2}\bigg(a_c^\dagger \sigma^-_1 + a_c \sigma^+_1 + a_c^\dagger \sigma^-_2 + a_c \sigma^+_2\bigg)
\end{eqnarray}

\begin{eqnarray}
    V &=& -\frac{g \tau^2}{48} \bigg( \Delta^2_1 \bigg(a_c^\dagger \sigma^-_1 + a_c \sigma^+_1 \bigg) + \Delta^2_2 \bigg(a_c^\dagger \sigma^-_2 + a_c \sigma^+_2 \bigg) \bigg) \\
    &\quad& -\frac{g^2 \tau^2}{24} \bigg(\Delta_1 \bigg(a_c^\dagger a_c + \frac{1}{2} \bigg)\sigma^z_1 + \Delta_2 \bigg(a_c^\dagger a_c + \frac{1}{2} \bigg)\sigma^z_2 \bigg)\\
    &\quad& -\frac{g^2 \tau^2}{48} \bigg( (\Delta_1 + \Delta_2)(\sigma^+_1 \sigma^-_2 + \sigma^-_1 \sigma^+_2) \bigg)
\end{eqnarray}

\noindent The eigenvalues and eigenstates of $H_0$ in the single excitation subspace $\{ \ket{e_1 0_c g_2}, \ket{g_1 1_c g_2}, \ket{g_1 0_c e_2} \}$, with $g_1=g_2=g/2$ are,

\begin{eqnarray}
    E^{(0)}_1 &=& 0\\
    E^{(0)}_2 &=& \dfrac{g}{\sqrt{2}}\\
    E^{(0)}_3 &=& -\dfrac{g}{\sqrt{2}}
    \label{eq:unperturbed_eigenvalues}
\end{eqnarray}

\begin{eqnarray}
    \ket{E^{(0)}_1} &=& \dfrac{1}{\sqrt{2}} \ket{e_1 0_c g_2} -  \dfrac{1}{\sqrt{2}} \ket{g_1 0_c e_2} \\
    \ket{E^{(0)}_2} &=& \dfrac{1}{2} \ket{e_1 0_c g_2} +  \dfrac{1}{2} \ket{g_1 0_c e_2} +  \dfrac{1}{\sqrt{2}} \ket{g_1 1_c g_2}\\
    \ket{E^{(0)}_2} &=& \dfrac{1}{2} \ket{e_1 0_c g_2} +  \dfrac{1}{2} \ket{g_1 0_c e_2} -  \dfrac{1}{\sqrt{2}} \ket{g_1 1_c g_2}\\
\end{eqnarray}

\noindent The first order corrections to the energies, $E_n^{(1)} = \langle E_n^{(0)} | V | E_n^{(0)} \rangle$ are,

\begin{eqnarray}
    E^{(1)}_1 &=& \langle E^{(0)}_1 | V | E^{(0)}_1 \rangle = \dfrac{g^2 \tau^2}{48} (\Delta_1 + \Delta_2 ) \\
    E^{(1)}_2 &=& \langle E^{(0)}_2 | V | E^{(0)}_2 \rangle = \dfrac{g^2 \tau^2}{48} (\Delta_1 + \Delta_2 ) 
                - \dfrac{g \tau^2}{48 \sqrt{2}} (\Delta_1^2 + \Delta_2^2 ) \\
    E^{(1)}_3 &=& \langle E^{(0)}_3 | V | E^{(0)}_3 \rangle = \dfrac{g^2 \tau^2}{48} (\Delta_1 + \Delta_2 ) 
                + \dfrac{g \tau^2}{48 \sqrt{2}} (\Delta_1^2 + \Delta_2^2 ) \\
\end{eqnarray}

\noindent The first order corrections to the energy eigenstates, $| E^{(1)}_n \rangle = \sum_{k \ne n} \dfrac{\langle E_k^{(0)} | V | E_n^{(0)} \rangle}{E_n - E_k} | E^{(0)}_k \rangle $ are,

\begin{eqnarray}
    | E^{(1)}_1 \rangle &=& \bigg( \dfrac{\tau^2}{48 \sqrt{2}} (\Delta_1^2 - \Delta_2^2) 
                                    + \dfrac{g \tau^2}{48} (\Delta_1 - \Delta_2) \bigg) | E^{(0)}_2 \rangle
                        + \bigg( \dfrac{\tau^2}{48 \sqrt{2}} (\Delta_1^2 - \Delta_2^2) 
                                    - \dfrac{g \tau^2}{48} (\Delta_1 - \Delta_2) \bigg) | E^{(0)}_3 \rangle \\
    | E^{(1)}_2 \rangle &=& - \bigg( \dfrac{\tau^2}{48 \sqrt{2}} (\Delta_1^2 - \Delta_2^2) 
                                    + \dfrac{g \tau^2}{48} (\Delta_1 - \Delta_2) \bigg) | E^{(0)}_1 \rangle
                                    - \dfrac{g \tau^2}{24 \sqrt{2}} (\Delta_1 + \Delta_2) \bigg) | E^{(0)}_3 \rangle \\
    | E^{(1)}_3 \rangle &=& - \bigg( \dfrac{\tau^2}{48 \sqrt{2}} (\Delta_1^2 - \Delta_2^2) 
                                    - \dfrac{g \tau^2}{48} (\Delta_1 - \Delta_2) \bigg) | E^{(0)}_1 \rangle
                                    + \dfrac{g \tau^2}{24 \sqrt{2}} (\Delta_1 + \Delta_2) \bigg) | E^{(0)}_2 \rangle
\end{eqnarray}

\noindent In the unperturbed eigenbasis, we have,

\begin{eqnarray}
    | \Psi(0) \rangle &=& |e_1 0_c g_2 \rangle = \dfrac{1}{\sqrt{2}} | E^{(0)}_1 \rangle
                            + \dfrac{1}{2} | E^{(0)}_2 \rangle + \dfrac{1}{2} | E^{(0)}_3 \rangle \\
    | \Psi_{\mathrm{target}} \rangle &=& |g_1 0_c e_2 \rangle = -\dfrac{1}{\sqrt{2}} | E^{(0)}_1 \rangle
                            + \dfrac{1}{2} | E^{(0)}_2 \rangle + \dfrac{1}{2} | E^{(0)}_3 \rangle \\
\end{eqnarray}

\noindent Rearranging the perturbation relation, we get $|E_i^{(0)} \rangle = |E_i \rangle - \lambda | E^{(1)}_i \rangle$. To first order in $\lambda$, the unperturbed kets can be expressed purely in terms of the perturbed kets. In the perturbed eigenbasis, the fidelity of state transfer at time $t$ is given by,

\begin{eqnarray}
    F &=& |\langle \Psi_{target} | \Psi(t) \rangle|^2 \\
    &=& |\langle g_1 0_c e_2 | U(t) | e_1 0_c g_2 \rangle |^2 \\
    &=& |\langle g_1 0_c e_2 | e^{-\mathrm{i}\bar{H}t} | e_1 0_c g_2 \rangle |^2 \\
    &=& \bigg| - \bigg(\dfrac{1}{2} + \dfrac{\tau^2}{48} (\Delta_1^2 - \Delta_2^2) \bigg) \mathrm{e}^{-\mathrm{i}E_1 t} \\
    &\quad\quad+& \bigg( \dfrac{1}{4} - \tau^2 \bigg( \dfrac{\Delta_1^2 - \Delta_2^2}{96} + \dfrac{g}{48 \sqrt{2}} (\Delta_1 - \Delta_2 ) + \dfrac{g}{48 \sqrt{2}} (\Delta_1 + \Delta_2 ) \bigg) \mathrm{e}^{-\mathrm{i}E_2 t}  \\
    &\quad\quad+& \bigg( \dfrac{1}{4} - \tau^2 \bigg( \dfrac{\Delta_1^2 - \Delta_2^2}{96} - \dfrac{g}{48 \sqrt{2}} (\Delta_1 - \Delta_2 ) - \dfrac{g}{48 \sqrt{2}} (\Delta_1 + \Delta_2 ) \bigg) \mathrm{e}^{-\mathrm{i}E_3 t} 
    \bigg|^2
\end{eqnarray}

\noindent The above expression can be simplified using 
\begin{eqnarray}
    \bigg| c_1 \mathrm{e}^{-\mathrm{i} E_1 t} + c_2 \mathrm{e}^{-\mathrm{i} E_1 t} + c_3 \mathrm{e}^{-\mathrm{i} E_1 t} \bigg|^2 
    &=& c_1^2 + c_2^2 + c_3^2 + 2 c_1 c_2
    + \cos[(E_1 - E_2)t] \\
    &\quad\quad+& 2 c_1 c_3 \cos[(E_1 - E_3)t]
        + 2 c_2 c_3 \cos[(E_2 - E_3)t]
\end{eqnarray}

\noindent where $c_i$ are real.

\begin{eqnarray}
    F &=& \dfrac{6}{16} + \dfrac{\tau^2}{96} (\Delta_1^2 - \Delta_2^2 ) 
            - \dfrac{1}{2} \cos \bigg[ \bigg( \dfrac{g}{\sqrt{2}} - \dfrac{g \tau^2}{48 \sqrt{2}} (\Delta_1^2 + \Delta_2^2) \bigg) t \bigg] \\
            &\quad\quad+& \bigg( \dfrac{1}{8} - \dfrac{\tau^2}{96} (\Delta_1^2 - \Delta_2^2 ) \bigg) \cos \bigg[ \bigg(
            \sqrt{2}g - \dfrac{g \tau^2}{24 \sqrt{2}} (\Delta_1^2 + \Delta_2^2) \bigg) t \bigg]
\end{eqnarray}

\noindent Using the identity $\cos(A - B) = \cos A \cos B + \sin A \sin B$, the small angle approximation $\cos \varepsilon \approx 1 - \dfrac{\varepsilon^2}{2}$, and setting $t=2t_f^0=\sqrt{2} \pi / g$, the expression for fidelity simplifies to

\begin{eqnarray}
    F \approx 1 - \dfrac{\pi^2 \tau^4}{4608} (\Delta_1^2+\Delta_2^2)^2
\end{eqnarray}

\noindent The infidelity $1-F$ is then

\begin{equation}
    1 - F  \approx \dfrac{\pi^2 \tau^4}{4608} (\Delta_1^2+\Delta_2^2)^2
\end{equation}

\section{Numerical methods}
\label{sec:Appendix_Numerical_Method}

\noindent In order to compute the fidelity $F(t)$ of state transfer, we solve for the time evolved state of the system $|\Psi(t)\rangle=U(t)|\Psi(0)\rangle$ governed by the POCEG Hamiltonian $H$ given in (\ref{eq:Hamiltonian_POCEG}) using  a step-wise diagonalization of the Hamiltonian by considering it to be piece-wise time-independent. The time evolution from $t=0$ to $t=2t_f^0$ is broken into small steps $\delta t$ over which the Hamiltonian is assumed to be time-independent. The evolution operator for each interval $t \rightarrow t + \delta t$ is calculated as $U(t+\delta t, t)=\exp(-\mathrm{i}Ht)$. Following the Carr-Purcell sequence, $\pi_x$ pulses are applied instantaneously at $t=(n - 1/2) \tau$ for $n=1, 2, 3, \cdot\cdot\cdot$, using $\sigma_x$ rotations on both qubits simultaneously. After every odd numbered pulse, the qubit-cavity couplings are turned off ($g_1=g_2=0$). After $t=2t_f^0$, the fidelity of the operation is calculated as $F(t) = |\langle \Psi_{\mathrm{target}}|\Psi(t)\rangle|^2$. \\

\noindent For the results in Fig.~\ref{fig:Noisy_Protocol_Performance} which show the performance of POCEG when the detunings fluctuate over time at various rates, the Hamiltonian $H(t)$ is updated with instantaneous detuning values $\Delta_{1,2}(t)$ for each time evolution step $t \rightarrow t + \delta t$. These detunings are sampled from their corresponding Gaussian distribution centered at $\Delta_{1,2}^0$ with standard deviation $\sigma$ and correlation time $\tau_c$. The results represent data collected over $100$ runs. \\

\noindent We treat the case of cavity damping, the results for which are shown in Fig. ~\ref{fig:Damped_Protocol_Performance}, using the master equation given by \cite{Cohen_Tannoudji_Book1992},
\begin{eqnarray}
    \dfrac{d}{dt} \rho_s = -\dfrac{\mathrm{i}}{\hbar} [H, \rho_s] - \dfrac{\kappa}{2} (a^{\dagger} a \rho_s + \rho_s a^{\dagger} a)
                            + \kappa a \rho_s a^{\dagger}
\end{eqnarray}
\noindent where $\rho_s$ is the density matrix of the \textit{Qubit 1}-\textit{Cavity}-\textit{Qubit 2} system, and $\kappa$ is the damping rate of the cavity. The time evolution of the density matrix is performed in a manner similar to that described above.\\
 
\noindent In the results presented in this paper, we use $\omega_c=600g$. The parameter values used are, in general, in line with recent experimental work with photon-mediated two-qubit gates \cite{PhysRevX.12.021026, DijkemaVandersypen_NatPhys2025}. The detuning range we consider is, however, several times higher, chosen as such to demonstrate the robustness of the protocol.

\section{Performance of POCEG for transfer of the equal superposition state}
\label{sec:Appendix_Superposition_transfer}

\begin{figure}[t] 
    \centering
    \includegraphics[width=0.5\columnwidth]{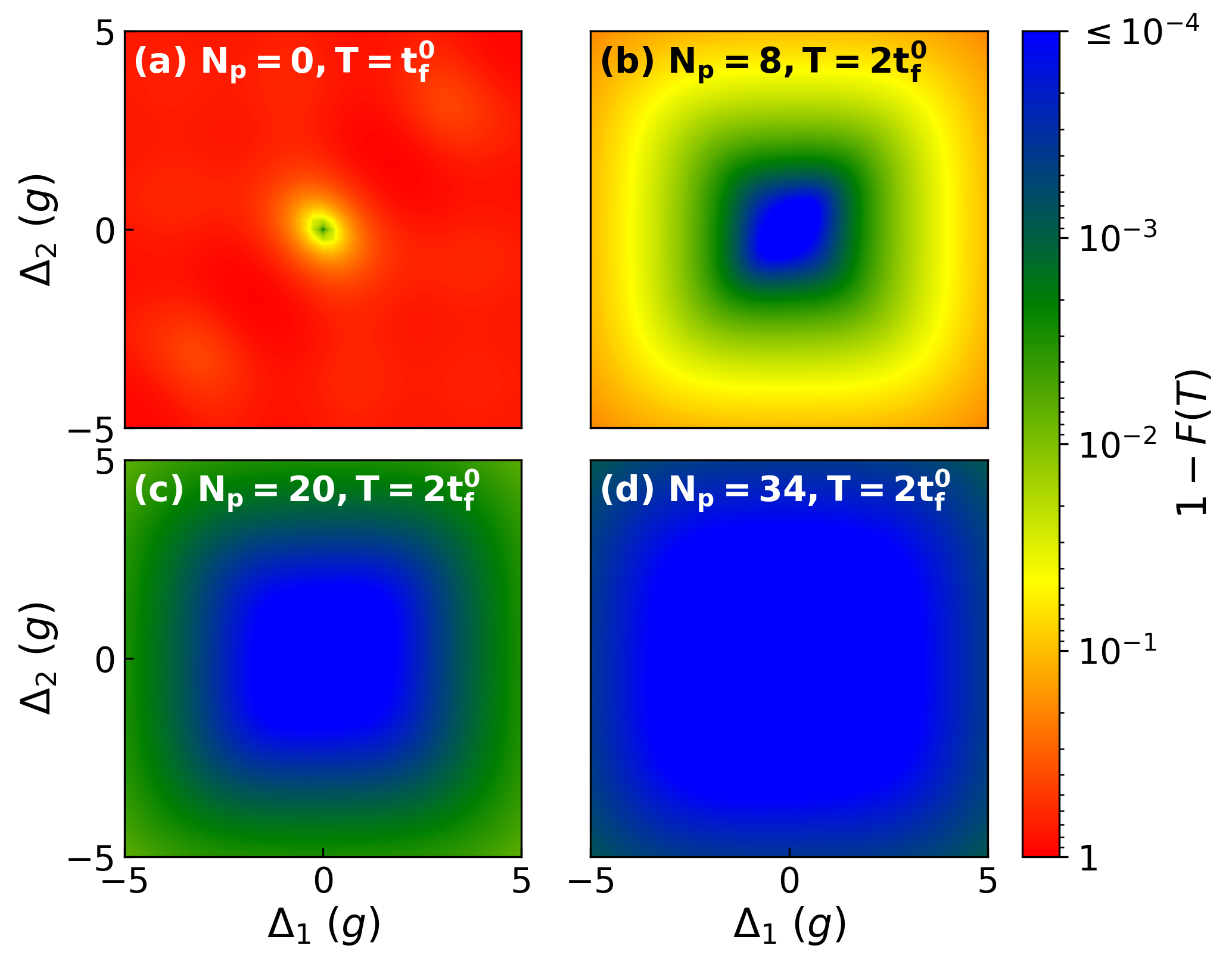}
    \caption{Performance POCEG protocol for the transfer of the equal superposition state measured as the infidelity $1-F$ as a function of all  detuning combinations ($\Delta_1,\Delta_2$) that represent the quasistatic regime for a noise bandwidth $10g$ centered at the cavity frequency. \textbf{(a):} Without POCEG. \textbf{(b, c, d):} With POCEG for different number of pulses $N_p$.} 
    \label{fig:POCEG_superposition}
\end{figure}

\noindent We discuss the performance of POCEG for the transfer of the state $|\Phi^l_i\rangle = \dfrac{1}{2} |g_1\rangle + \dfrac{1}{2} |e_1 \rangle$ from \textit{Qubit 1} to \textit{Qubit 2} via the cavity. Here, $|\Psi(t=0)\rangle=\dfrac{1}{2} |g_1 0_c g_2\rangle + \dfrac{1}{2}|e_1 0_c g_2\rangle $ and $ |\Psi_{\mathrm{target}}\rangle=\dfrac{1}{2} |g_1 0_c g_2\rangle + \dfrac{1}{2}|g_1 0_c e_2\rangle $. The fidelity of state transfer after time $t > 0$ is $F(t) = |\langle \Psi_{\mathrm{target}}|\Psi(t)\rangle|^2$. Figure \ref{fig:POCEG_superposition} shows the fidelity for any combination of detunings with $-5g \le \Delta_1, \; \Delta_2 \le 5g$. Panel (a) illustrates the uncontrolled system for which only a small region around $(\Delta_1 = 0, \Delta_2 = 0)$ shows optimal fidelity. Panels (b), (c) and (d) indicate that the region of optimal fidelity gradually increases with the number of pulses. We note that here, fewer pulses are required to achieve a certain target fidelity compared to the fully polarized case. This is because the ground state component $|g_1 0_c g_2 \rangle$ is largely a dark state in the evolution dynamics since the only terms in the Hamiltonian that are capable of engaging with them are the weak (for $g/\omega_c<<1$) energy non-conserving terms. Therefore, the $|g_1 0_c g_2 \rangle$ component is preserved during evolution and contributes to $\sim 50\%$ of the fidelity regardless of the detuning. This feature is discussed in a recent work that characterizes the performance of two-qubit gates over a broad parameter space \cite{yuvarajan2025cavity}.

\section{Occupation probability of higher excited states of the cavity and its effect on the state transfer fidelity}
\label{sec:Appendix_Cavity_Cutoff}

\noindent A study of the probability of occupation of higher excited states $(n_c>1)$ of the cavity is necessary in order to identify a reasonable cutoff at which the cavity levels can be truncated in the numerical simulations. Higher cavity levels can be excited due to the energy non-conserving terms in the Hamiltonian (\ref{eq:Hamiltonian_POCEG}) or due to the application of $\pi$ pulses to the qubits. It has been shown that the effect of the energy non-conserving terms scales roughly as $(g/\omega_c)^{2 \lfloor n_c/2 \rfloor}$ \cite{yuvarajan2025cavity}, where $n_c$ is the cavity level. Therefore, with $\omega_c>>g$, one can suppress the effect of these terms. The simulations in the paper use $\omega_c=600g$. On the other hand, $\pi$ pulses can produce higher excitations in the cavity in two scenarios- one when they are applied with the qubits coupled to the cavity, and the other when the qubits remain coupled to the cavity after an odd number of pulses. However, these concerns are addressed by our protocol where the instantaneous qubit-cavity couplings are turned off during the application of the $\pi$ pulses and after the application of an odd number of $\pi$ pulses.\\

\begin{figure}[t] 
    \centering
    \includegraphics[width=0.5\columnwidth]{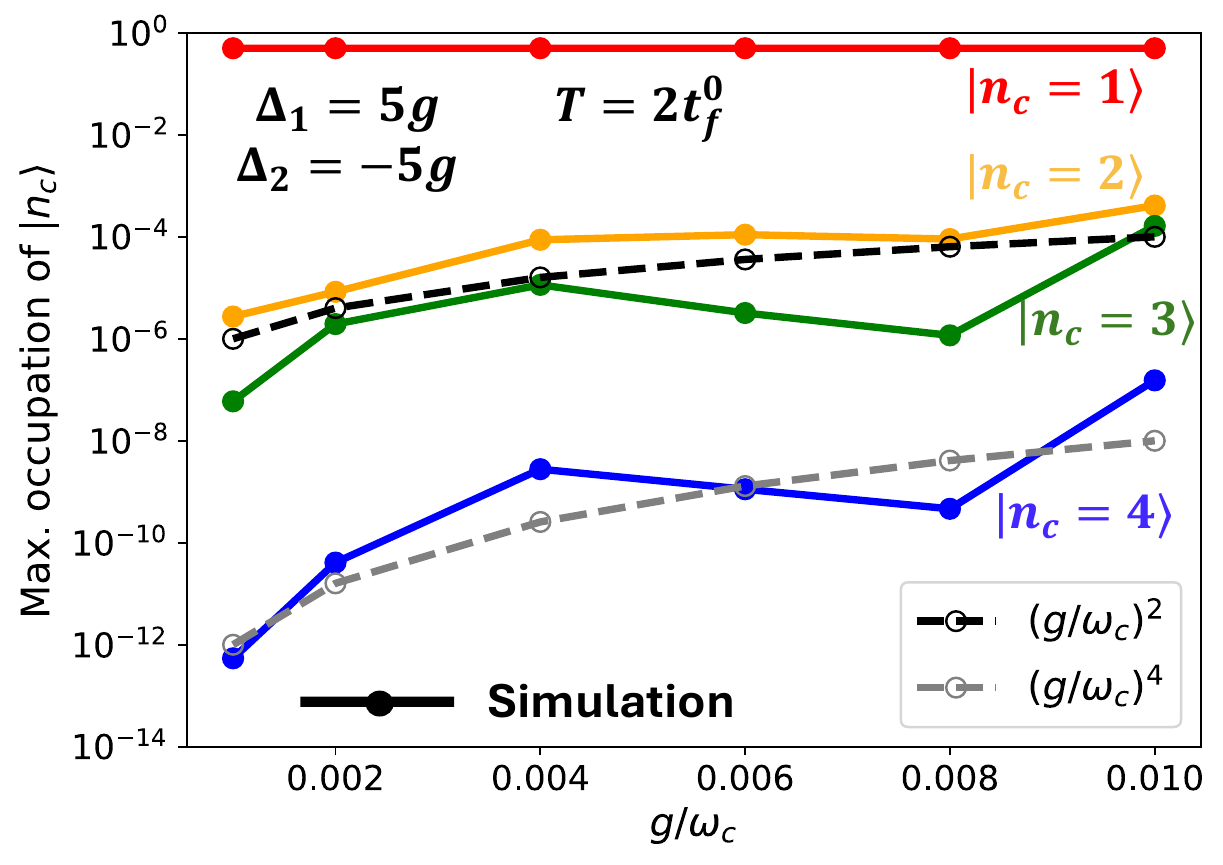}
    \caption{Maximum cavity occupation during time $T=2 tf_0$ as a function of the coupling $g/ \omega_c$ for $\Delta_1=-\Delta_2=\Delta=5g$ and number of pulses $N_p=42$. Solid lines represent simulation results and the dashed lines represent analytical esimates that scale as $(g/\omega_c)^{2 \lfloor n/2 \rfloor}$.}
    \label{fig:Cavity_Occupation_Pulsed}
\end{figure}

\noindent As shown in figure \ref{fig:Cavity_Occupation_Pulsed}, we find that the cavity can be truncated to 3 levels $(n^{\mathrm{max}}_c=3)$ to reasonably capture the state transfer fidelity to a precision of $10^{-4}$. For $n^{\mathrm{max}}_c=4$, the cavity occupation is relatively negligible, of the order $10^{-10}$ for $\omega_c=600g$, and will be even lower for higher cavity levels due to the scaling relation mentioned above. 

\end{document}